\documentclass[conference]{IEEEtran}

\usepackage{amsmath,amssymb}
\usepackage{graphicx}
\usepackage{booktabs}
\usepackage{hyperref}
\usepackage{cite}
\usepackage{algorithm}
\usepackage{algorithmic}
\usepackage{multirow}
\usepackage{array}
\usepackage{xcolor}
\usepackage{makecell}

\hypersetup{
    colorlinks=true,
    linkcolor=blue,
    citecolor=blue,
    urlcolor=blue
}

\title{Bangla-WhisperDiar: Fine-Tuning Whisper and PyAnnote for Bangla Long-Form Speech Recognition and Speaker Diarization}

\author{
Mohammed Aman Bhuiyan, Md Sazzad Hossain Adib, Samiul Basir Bhuiyan, Amit Chakraborty,\\
Aritra Islam Saswato, Ahmed Faizul Haque Dhrubo, Mohammad Ashrafuzzaman Khan\\
\textit{Department of ECE, North South University, Dhaka, Bangladesh}\\
{\small \textit{\{mohammed.aman, sazzad.adib, samiul.bhuiyan, chakraborty.amit,\}}}\\
{\small \textit{\{aritra.saswato, ahmed.dhrubo, mohammad.khan02\}@northsouth.edu}}
}

\begin{document}

\maketitle

\begin{abstract}
Automatic Speech Recognition (ASR) and speaker diarization in Bangla remain challenging due to long-form recordings, diverse acoustic conditions, and significant speaker variability. This work addresses these two core tasks in Bangla spoken language understanding by developing robust systems for long-form ASR and speaker diarization. For ASR (Problem 1), we fine-tune the \texttt{tugstugi\_bengaliai-regional-asr\_whisper-medium} model on a custom-curated dataset of approximately 15,000 chunked and aligned Bangla audio segments, employing full-weight training with extensive data augmentation including noise injection, reverb simulation, echo, clipping distortion, and pitch/time perturbation. For speaker diarization (Problem 2), we fine-tune the \texttt{pyannote/segmentation-3.0} model using PyTorch Lightning on the competition's annotated diarization dataset, swapping the fine-tuned segmentation backbone into the \texttt{pyannote/speaker-diarization-community-1} pipeline while retaining the pretrained speaker embedding and clustering components. Our ASR system achieves a Word Error Rate (WER) of \textbf{24.41}\%, while our diarization system achieves a Diarization Error Rate (DER) of \textbf{23.92}\%, both evaluated on the test set, demonstrating notable improvements over the respective pretrained baselines. We describe our complete pipeline, including data preprocessing, text normalization, audio augmentation, training strategies, inference optimization, and post-processing for both tasks. The corresponding code has been made publicly available at 
\url{https://github.com/sazzadadib/BitwiseMind_DL_Sprint_4.0}.
\end{abstract}


\section{Introduction}

Bangla (Bengali), spoken by over 230 million people worldwide, is one of the most widely spoken languages globally, yet it remains significantly under-resourced in the domain of speech technology. Long-form audio understanding, which includes both accurate transcription and speaker attribution, is critical for applications ranging from media indexing and meeting summarization to accessibility tools and legal or forensic audio analysis~\cite{radford2023whisper}.

Despite major progress in end to end ASR and large pretrained models, most state of the art systems remain optimized for high resource languages like English, resulting in higher error rates for Bangla, especially in spontaneous, multi speaker, and diverse domain audio. Addressing this gap requires carefully curated datasets, robust modeling approaches, and evaluation strategies tailored to the unique linguistic and acoustic characteristics of Bangla.
DL Sprint 4.0 poses two challenging problems on Bangla long-form audio:

\begin{figure}[t]
\centering
\includegraphics[width=\columnwidth]{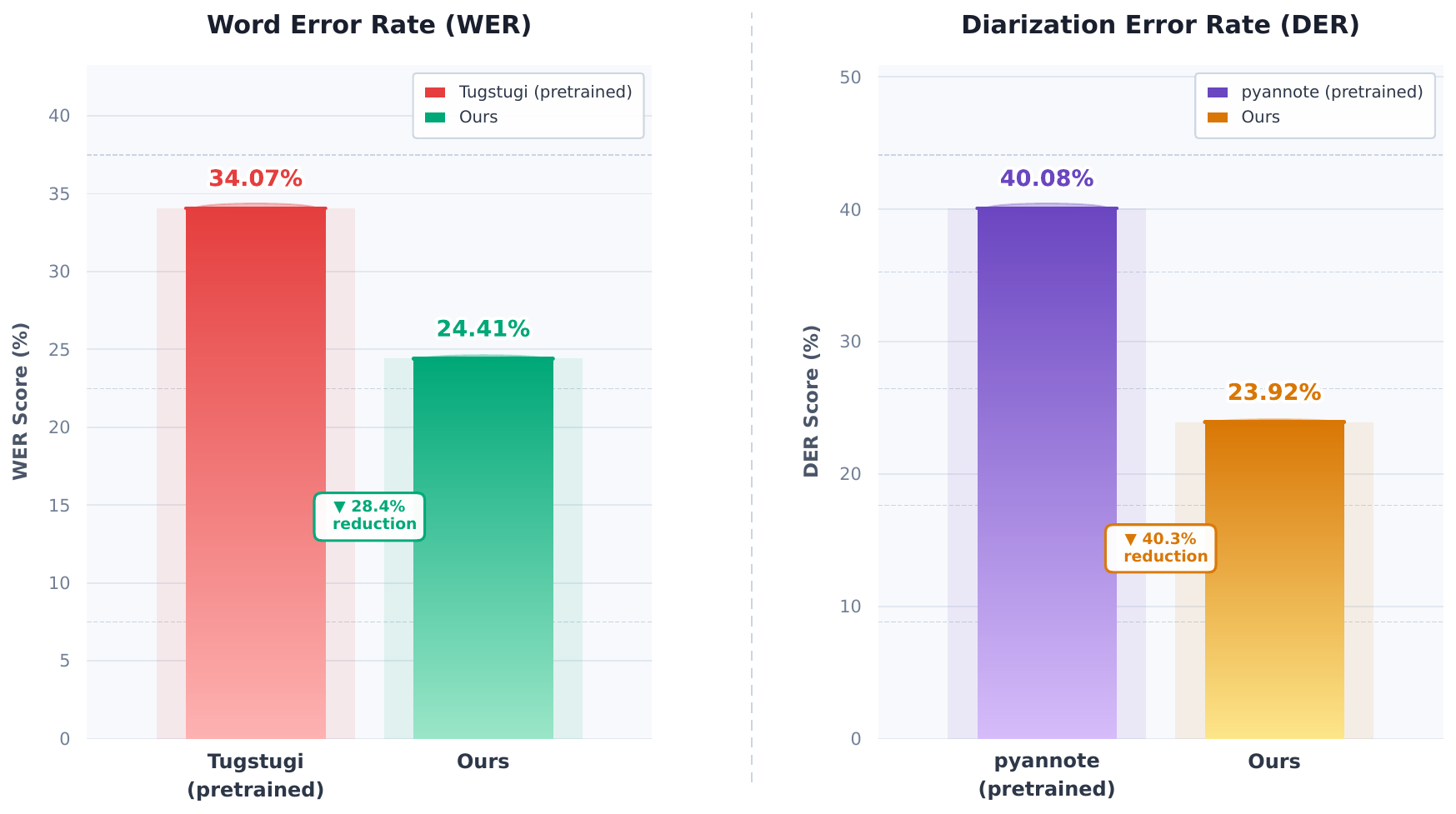} 
\caption{Comparison of pretrained baselines and the proposed system. Our method improves both WER and DER.}
\label{fig1}
\end{figure}
\begin{itemize}
    \item \textbf{Problem 1 — Bangla Long-Form ASR:} Transcribe extended Bangla speech recordings into text, evaluated using Word Error Rate (WER).
    \item \textbf{Problem 2 — Speaker Diarization:} Determine ``who spoke when'' by producing speaker-labelled time segments, evaluated using Diarization Error Rate (DER).
\end{itemize}

Both problems present unique challenges for Bangla. The ASR task must handle dialectal variation, conversational speech, environmental noise, and the morphological complexity of the Bangla script. The diarization task must accurately segment long recordings with potentially overlapping speakers and varying acoustic conditions. In this paper, we present our approach to both problems. For ASR, we leverage OpenAI’s Whisper architecture~\cite{radford2023whisper}, specifically the medium variant pre-trained on Bengali by Tugstugi~\cite{Tugstugi}, and perform  fine-tuning with a comprehensive data augmentation pipeline. For diarization, we adopt PyAnnote.Audio’s segmentation-3.0 model~\cite{plaquet2023pyannote, bredin2023pyannote} and fine-tune it on the competition dataset, integrating the fine-tuned weights into the full diarization pipeline.
\vspace{1.1cm}
\\
\noindent\textbf{Our key contributions are:}
\begin{enumerate}
    \item A complete data preprocessing pipeline for Bangla long-form audio, including automated chunking with ASR-guided ground-truth alignment using fuzzy matching.
    \item A comprehensive audio augmentation strategy combining noise injection, room reverb simulation, multi-tap echo, clipping distortion, bandpass filtering, pitch shifting, and time stretching to improve model robustness. Bangla-specific text normalization including English numeral-to-Bangla word conversion and non-Bengali character removal.
    \item Longform audio ASR model fine-tuning, followed by domain-adaptive fine-tuning of PyAnnote's segmentation model using checkpoint-based training and segmentation-swap inference for speaker diarization.
    \item Post-processing techniques including hallucination removal for ASR and short-segment filtering for diarization.
\end{enumerate}

\begin{figure*}[t]
    \centering
    \includegraphics[width=\textwidth]{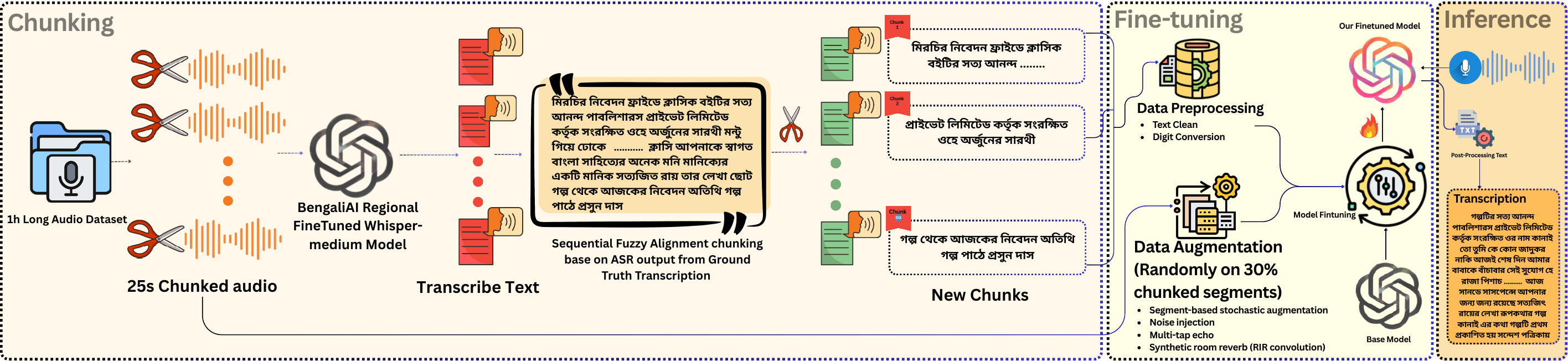}
    \caption{End-to-end pipeline for  Bangla Long-Form Speech Recognition, using fine-tuned Whisper-medium(BengaliAI) ASR model. The framework demonstrates data preprocessing, dataset chunking, model fine-tuning, and inference stages for transcribing long-form speech to standard Bangla transcription orthography.}
    \label{fig:figure2}
\end{figure*}
\section{Related Work}

\subsection{Automatic Speech Recognition}

Modern ASR systems have shifted from traditional Hidden Markov Model (HMM) and Connectionist Temporal Classification (CTC) approaches toward end-to-end encoder-decoder architectures. OpenAI's Whisper~\cite{radford2023whisper} represents a significant advance in multilingual ASR, trained on 680,000 hours of weakly supervised data spanning 99 languages. Whisper uses a Transformer-based encoder-decoder architecture that converts log-Mel spectrograms to text in an autoregressive fashion. While Whisper supports Bangla, its performance on long-form conversational Bangla speech remains suboptimal without domain-specific adaptation.

BengaliAI has released several fine-tuned Whisper variants for regional Bengali ASR~\cite{bengaliai2024whisper}, which serve as strong starting points for further fine-tuning. Data augmentation for speech has been extensively studied, including SpecAugment~\cite{park2019specaugment}, noise injection, and room impulse response (RIR) simulation~\cite{ko2017study}, all of which we adapt for our pipeline.

\subsection{Speaker Diarization}

Speaker diarization has evolved from clustering-based methods (e.g., agglomerative hierarchical clustering on speaker embeddings) to end-to-end neural approaches. PyAnnote.Audio~\cite{bredin2023pyannote, plaquet2023pyannote} provides a modular pipeline that combines neural speaker segmentation with speaker embedding extraction and clustering. The segmentation-3.0 model uses a PyanNet architecture to produce frame-level speaker activity probabilities; these are then paired with speaker embeddings to perform full diarization.

Recent work on low-resource diarization has shown that fine-tuning pretrained segmentation models on target-domain data can significantly reduce DER~\cite{plaquet2023pyannote}. Our approach follows this paradigm, adapting the multilingual pretrained model to the specific acoustic and speaker characteristics of Bangla conversational audio.

\section{Methodology}
\subsection{Problem 1 — Bangla Long-Form ASR}

Our ASR pipeline in Figure \ref{fig:figure2} consists of four main stages: (1) data preprocessing and chunking, (2) text normalization and cleaning, (3) model fine-tuning, and (4) inference with post-processing.

\subsubsection{Data Preprocessing and Audio Chunking}

Long-form audio recordings (varying from several minutes to over an hour) are first chunked into segments of at most 25 seconds, the practical input limit for Whisper's encoder. We employ a fixed-length non-overlapping chunking strategy:

\begin{enumerate}
    \item Each training audio file is loaded at 16\,kHz.
    \item The audio is divided into non-overlapping 25-second segments.
    \item Silent or near-silent trailing fragments (below 0.5\,s) are discarded.
    \item Each chunk is transcribed using the pretrained BengaliAI Whisper model to obtain an ASR hypothesis.
\end{enumerate}

To obtain ground-truth labels for each chunk, we implement a \textbf{sequential fuzzy alignment} algorithm. The full ground-truth transcript is split into words, and a sequential pointer is maintained across the corpus. For each ASR-transcribed chunk, a bidirectional search window ($\pm 5$ words) and span-length variation ($\pm 3$ words) are used to find the best-matching ground-truth span via \texttt{RapidFuzz} ratio scoring. This ensures reading-order preservation while accommodating minor ASR insertions and deletions.

\subsubsection{Text Normalization and Cleaning}

We apply several Bangla-specific normalization steps to ensure consistency and linguistic correctness in the transcripts. First, all English-digit sequences are converted into their Bangla word equivalents using the \texttt{num2words} library, with four-digit numbers in the range 1000–2099 specifically interpreted as calendar years. Next, characters outside the Bengali Unicode block (U+0980–U+09FF) are removed, while allowing digits, whitespace, and common punctuation marks. Finally, multiple consecutive whitespace characters are collapsed into a single space to maintain clean and uniform text formatting.

\subsubsection{Audio Data Augmentation}

To improve robustness to real-world acoustic conditions, we apply a stochastic augmentation pipeline to approximately 30\% of each training audio clip’s duration. Augmentations are performed on non-overlapping random windows of 3–6 seconds. Within these windows, pink and white noise are mixed and added with a 65\% probability using a randomly selected scaling factor. A multi-tap echo effect is applied with 55\% probability, introducing 2–4 echo taps with random delays between 150–800,ms and decay factors ranging from 0.4 to 0.75. Room reverberation is incorporated with 60\% probability through convolution with a synthetic Room Impulse Response (RIR), generated for randomly selected room sizes (small, medium, or large) and configurable RT60 decay values. Additionally, analog-to-digital clipping distortion is simulated with 30\% probability. Bandpass filtering is applied with 20\% probability to simulate telephone-band audio (300–3400,Hz). Pitch shifting of up to $\pm 3$ semitones is performed with 25\% probability using \texttt{librosa}, and time stretching is applied with 25\% probability using a rate factor sampled between 0.80 and 1.20. Each augmented segment is individually peak-normalized before being reinserted into the original audio stream.

\subsubsection{Model Architecture and Fine-Tuning}

We use the Tugstugi~\cite{Tugstugi} model as our base, which is built upon Whisper-Medium ($\sim$764M parameters) and already fine-tuned on Bengali regional speech data. To maximize adaptation capacity, we performed fine-tuning with loaded weights rather than parameter-efficient adaptation.

Training is conducted using the AdamW 8-bit optimizer adamw-bnb-8bit, which reduces VRAM usage by approximately 2.4 GB compared to standard fp32 AdamW while maintaining equivalent convergence behavior. We employ fp16 mixed precision training with a learning rate of $5 \times 10^{-6}$, scheduled using a cosine decay with 100 warmup steps. The batch size is set to 8 per GPU with 2 gradient accumulation steps, resulting in an effective batch size of 16. The model is trained for 6 epochs, with total steps computed dynamically based on dataset size. Input constraints include a maximum audio duration of 30s and a maximum transcript length of 1000 tokens. Gradient checkpointing is enabled to further reduce VRAM consumption. The best model checkpoint is selected based on the lowest Word Error Rate (WER) on the validation set.
Audio inputs are converted into 80-channel log-Mel spectrograms using \texttt{WhisperFeatureExtractor}, and transcripts are tokenized with the Bangla Whisper tokenizer.

\subsubsection{Inference Pipeline}

During inference, test audio files are first loaded and resampled to 16,kHz mono. Each audio sample is then segmented into 25-second chunks, with zero-padding applied to the final chunk when necessary to maintain consistent input length. The segments are processed in batches of 8 using beam search decoding with num-beams=4 and max-length=448. To accelerate inference, static KV caching was enabled.

\noindent\textbf{Post-processing} includes NFC Unicode normalization and removal of zero-width characters to ensure clean text output. Hallucinated repetitions are mitigated through iterative regex-based deduplication targeting repeated multi-word sequences, single-word repetitions, and character-level n-gram patterns. Finally, speaker-change markers (e.g., \texttt{>>}) are removed from the generated transcripts.


\subsection{Problem 2 — Speaker Diarization}

\begin{figure}[t]
    \centering
    \includegraphics[width=\columnwidth]{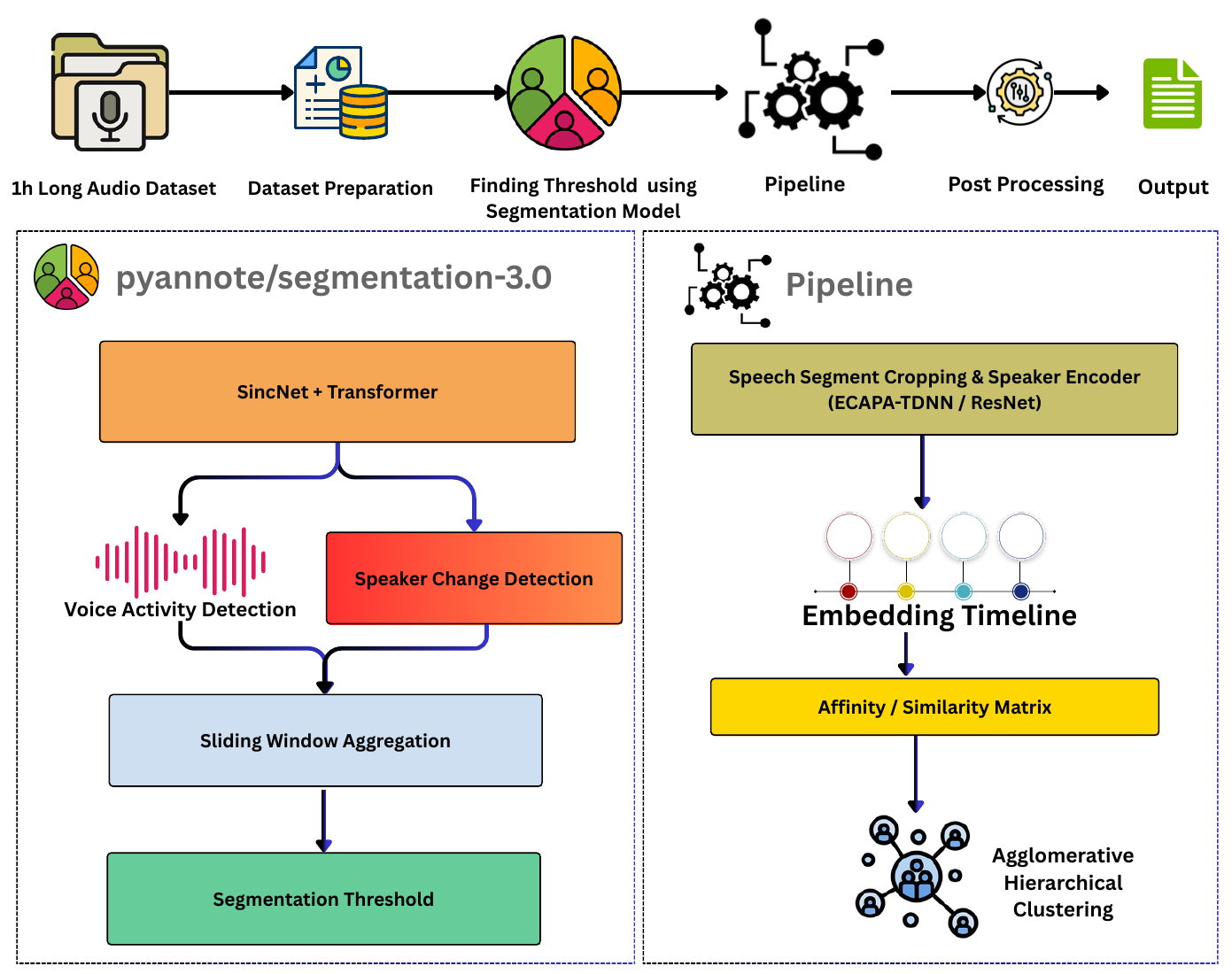}
    \caption{Complete end-to-end pipeline for fine-tuned Bengali speaker diarization using pyannote.audio 4.0. The segmentation model (SincNet + Transformer Encoder) is fine-tuned on DL Sprint 4.0 Bengali data while the speaker embedding and clustering modules remain frozen at pretrained weights}
    \label{fig:figure3}
\end{figure}

Our diarization pipeline in Figure \ref{fig:figure3} consists of (1) data preparation, (2) model fine-tuning, (3) pipeline construction, and (4) post-processing.

\subsubsection{Data Preparation}

The competition provides annotated training files with CSV annotations containing start-time, end-time, and speaker-id fields in HH:MM:SS format. We convert these annotations into the standard formats required by PyAnnote. Specifically, we generate RTTM  files to encode speaker turn segments with their corresponding start times and durations. We also create UEM  files to define the full scoring region for each audio recording, spanning from time 0 to the total duration. In addition, LST files are prepared to enumerate file URIs for the train, development, and test splits, and a database.yml configuration file is constructed to enable PyAnnote’s protocol system to correctly locate the associated audio, RTTM, and UEM files. For validation, the last two files in the dataset are reserved for development, while the remaining files are used for training.

\subsubsection{Model Architecture}

The diarization system is built upon the modular pipeline provided by PyAnnote, which comprises three primary components. First, the \textbf{Segmentation} stage uses pyannote/segmentation-3.0, a PyanNet-based neural model that generates frame-level speaker activity probabilities. It operates on 10-second audio chunks and outputs per-frame speaker presence indicators, supporting up to three simultaneous speakers. Second, the \textbf{Speaker Embedding} component employs a pretrained speaker embedding model to extract fixed-dimensional speaker representations from the detected speech segments. Finally, the \textbf{Clustering} stage applies agglomerative clustering to group segments by speaker identity across the entire recording, thereby producing the final diarization output.

\subsubsection{Fine-Tuning}

We fine-tune only the segmentation component of the PyAnnote pipeline, while keeping the speaker embedding and clustering components at their pretrained weights. Training is conducted using the PyTorch Lightning framework\cite{pytorch_lightning} on single-GPU hardware (NVIDIA Tesla T4). The task is defined as \texttt{SpeakerDiarization} with 10 second audio chunks, a maximum of 3 speakers per chunk, and at most 1 speaker per frame. The model is trained for 50 epochs, with checkpointing configured to save the best model based on minimum training loss (\texttt{loss/train}), retaining both the single best checkpoint and the last checkpoint for reference.

\subsubsection{Pipeline Construction}

At inference time, the complete diarization pipeline is assembled by first loading the pretrained \texttt{pyannote/speaker-diarization-community-1} pipeline. The best fine-tuned segmentation checkpoint (e.g., \texttt{segmentation-epoch=39.ckpt}) is then loaded, and the segmentation \texttt{Inference} object within the pipeline is swapped with the fine-tuned model. All original pipeline parameters, including segment duration, step size, aggregation settings, and batch size, are preserved. This integration ensures that the fine-tuned segmentation model operates within the established and validated pipeline infrastructure.

\subsubsection{Post-Processing}

To improve output quality, we apply a minimum-duration segment filter with a 0.3 second threshold, removing spurious ultra-short segments that are likely artifacts of the segmentation model. The final diarization output is formatted as JSON, with each segment represented by start-time, end-time, and speaker-id fields.
\section{Experimental Setup}

\subsection{Dataset}

\subsubsection{ASR Dataset}

The ASR dataset is derived from the \textbf{BengaliLoop}\cite{BengaliLoop}  long-form benchmark (DL Sprint 4.0). The corpus contains \textbf{191 recordings} totaling \textbf{158.6 hours} and approximately \textbf{792k words}, collected from 11 public YouTube channels, primarily Bangla drama (natok), along with news and entertainment content.

Transcripts were obtained through subtitle extraction followed by \textit{human-in-the-loop} verification and correction to ensure alignment with spoken audio. Audio was standardized to 16 kHz mono WAV format. 

For our experiments:
\begin{itemize}
    \item Long recordings were segmented into $\leq$ 25-second chunks.
    \item Ground-truth transcripts were aligned to each segment.
    \item The final processed dataset contains approximately \textbf{15,229} samples.
    \item Train/validation split: \textbf{90\%/10\%}.
\end{itemize}

\subsubsection{Diarization Dataset}

The diarization dataset also is taken from the \textbf{BengaliLoop}\cite{BengaliLoop} speaker diarization benchmark, consisting of \textbf{24 manually annotated recordings} totaling \textbf{22 hours} with \textbf{5,744 speaker segments}.

The official split is:
\begin{itemize}
    \item \textbf{Train:} 10 recordings (9.5 hours, 2,612 segments).
    \item \textbf{Test:} 14 recordings (12.5 hours, 3,132 segments).
\end{itemize}

For our experiments:
\begin{itemize}
    \item Training/Validation split: All files except the last 2. Last 2 files used for evaluation during training.
\end{itemize}
Annotations are provided in CSV format with start time, end time, and speaker ID. Overlapping speech is assigned to the first speaker (single-label policy), following the benchmark protocol.

\subsection{Hardware}

All experiments are conducted on Kaggle's free-tier NVIDIA Tesla T4 GPU. 
Key memory management strategies include:

\begin{itemize}
    \item ASR: Gradient checkpointing, 8-bit AdamW optimizer, fp16 mixed precision.
    \item Diarization: PyTorch Lightning's built-in memory management with single-GPU training.
\end{itemize}

\subsection{Evaluation Metrics}

\subsubsection{Word Error Rate (WER)}
WER measures the edit distance between predicted and reference transcriptions at the word level:
\begin{equation}
    \text{WER} = \frac{S + D + I}{N}
\end{equation}
where $S$, $D$, $I$ denote the number of substitutions, deletions, and insertions, and $N$ is the total number of words in the reference. Lower WER is better.

\subsubsection{Diarization Error Rate (DER)}
DER is the standard metric for speaker diarization evaluation:
\begin{equation}
    \text{DER} = \frac{\text{FA} + \text{MISS} + \text{ERROR}}{\text{TOTAL}}
\end{equation}
where FA is false alarm (speech detected where none exists), MISS is missed speech, ERROR is speaker confusion, and TOTAL is the total reference speech duration. Lower DER is better.


\subsubsection{Real-Time Factor (RTF)}
Inference efficiency is measured as:
\begin{equation}
    \text{RTF} = \frac{T}{D}
\end{equation}
where $T$ is the total inference time and $D$ is the total audio duration. Lower RTF indicates faster inference.

\section{Results}

\subsection{Problem 1 — Bangla Long-Form ASR}

Table~\ref{tab:asr_results} compares different ASR models in terms of WER. The results show that the baseline HisabTitu-BN model performs poorly (50.67\%), while the pretrained Tugstugi model significantly reduces WER to 34.07\%. LoRA fine-tuning on Tugstugi further improves performance to 31.32\%. Our proposed fine-tuning approach achieves the best WER of 24.41\%, demonstrating a substantial relative improvement over all baselines and highlighting the effectiveness of our training and adaptation strategy.

\begin{table}[ht]
\centering
\caption{ASR Performance Comparison (WER \%)}
\label{tab:asr_results}
\begin{tabular}{lc}
\toprule
\textbf{Model} & \textbf{WER (\%)} \\
\midrule
1. HisabTitu-BN & 50.67 \\
2. Tugstugi (pretrained) & 34.07 \\
3. Tugstugi (LoRA fine-tune) & 31.32 \\
4. \textbf{Ours} & \textbf{24.41}$^{\star}$ \\
\bottomrule
\end{tabular}
\end{table}

Table~\ref{tab:post_processing} quantifies the impact of post-processing on ASR performance. Without post-processing, WER is 25.76\%, which is reduced to 24.41\% when post-processing is applied. This confirms that post-processing contributes a measurable improvement, complementing the gains achieved through model fine-tuning.

\begin{table}[ht]
\centering
\caption{Post-processing Effectiveness Analysis}
\label{tab:post_processing}
\begin{tabular}{lc}
\toprule
\textbf{Method} & \textbf{WER (\%)} \\
\midrule
1. Without Post-Processing & 25.76 \\
2. With Post-Processing & \textbf{24.41}$^{\star}$ \\
\bottomrule
\end{tabular}
\end{table}

\subsection{Problem 2 — Speaker Diarization}

Table~\ref{tab:der_results} reports Diarization Error Rate (DER) across different pipelines. Traditional VAD + clustering approaches perform poorly, with DERs of 73.71\% (WebRTC VAD) and 61.50\% (Silero VAD). The pretrained pyannote.audio model reduces DER to 40.08\%, while our fine-tuned system achieves 23.92\%, showing a dramatic improvement over both classical and pretrained methods. This emphasizes the effectiveness of our adaptation and speaker modeling approach.

\begin{table}[ht]
\centering
\caption{Speaker Diarization Performance (DER \%)}
\label{tab:der_results}
\begin{tabular}{lc}
\toprule
\textbf{Pipeline} & \textbf{DER (\%)} \\
\midrule
WebRTC VAD + ECAPA + Clustering & 73.71 \\
Silero VAD + ECAPA + Clustering & 61.50 \\
pyannote.audio (pretrained) & 40.08 \\
\textbf{Ours} & \textbf{23.92}$^{\star}$ \\
\bottomrule
\end{tabular}
\end{table}



\subsection{Inference Efficiency}

Table~\ref{tab:rtf} presents the Real-Time Factor (RTF) and inference time for the proposed system on NVIDIA T4 GPUs. Using a single GPU, the long-form ASR system processes a 3-hour 38-minute recording with an RTF of \textbf{0.1659}, meaning the system runs significantly faster than real time. With the proposed inference optimization and dual-GPU parallelization, the processing time is further reduced to \textbf{25 minutes}, achieving an RTF of \textbf{0.0190}, which demonstrates substantial acceleration.
It employs \textbf{CTranslate2-optimized} Whisper models with FP16 precision, reducing GPU memory usage while increasing inference speed. 

For the speaker diarization module, the system processes the same data in 1 hour 20 minutes with an RTF of \textbf{0.1054} on a single T4 GPU. These results indicate that both the ASR and diarization components are computationally efficient and capable of real-time or faster-than-real-time processing, making the pipeline suitable for large-scale or long-duration speech analysis tasks.

\begin{table}[ht]
\centering
\caption{Inference Efficiency (Real-Time Duration)}
\label{tab:rtf}
\begin{tabular}{lccc}
\toprule
\textbf{Task} & \textbf{Hardware} & \textbf{Inference Time} & \textbf{RTF} \\
\midrule
Long Form ASR & T4x1 GPU & 3hr 38min & 0.1659 \\
\makecell[l]{Long Form ASR \\ \small (Inference Optimized)} & T4x2 GPU & \textcolor{red}{\textbf{25min*}} & 0.0190 \\
Speaker Diarization & T4x1 GPU & 1hr 20min & 0.1054 \\
\bottomrule
\end{tabular}
\end{table}

\subsection{Competition Scores}

Table~\ref{tab:competition_scores} summarizes our final competition scores on the public and private test sets. For long-form ASR, WER is 23.58\% (public) and 24.75\% (private), while for speaker diarization, DER is 18.52\% (public) and 26.13\% (private). These results show consistent competitive performance across both datasets, confirming the robustness and generalization of our models.

\begin{table}[ht]
\centering
\caption{Competition Score Summary}
\label{tab:competition_scores}
\begin{tabular}{lcc}
\toprule
\textbf{Task} & \textbf{Public (29\%)} & \textbf{Private (71\%)} \\
\midrule
Long Form ASR (WER \%) & 23.580  & 24.750   \\
Speaker Diarization (DER \%) & 18.518  & 26.133  \\
\bottomrule
\end{tabular}
\end{table}

\section{Analysis and Discussion}

\subsection{ASR Analysis}

\subsubsection{Impact of Data Augmentation}
The stochastic augmentation pipeline proved essential for bridging the domain gap between the pretrained model's training data (cleaner Bengali speech) and the competition's long-form conversational audio. By simulating realistic degradations (noise, reverb, echo), the model became significantly more robust to the varied acoustic conditions present in the test data. 

\subsubsection{Text Normalization Benefits}
The Bangla-specific text normalization pipeline (particularly numeral-to-word conversion) substantially improved label consistency. Without this normalization, the model would encounter the same concept represented inconsistently as both digit sequences and Bangla words, leading to ambiguous learning signals. 
\subsubsection{Chunk Alignment Quality}
The sequential fuzzy-matching approach for ground-truth alignment ensures that each 25-second audio chunk receives an accurately corresponding text label, maintaining reading order while tolerating ASR errors in the alignment signal. This is critical because misaligned labels directly degrade fine-tuning quality.

\subsubsection{Post-Processing Effectiveness}
Whisper models are known to hallucinate, generating repetitive phrases, especially on noisy or silent audio segments. Our iterative regex-based deduplication removes three hallucination patterns: multi-word repeated phrases, single-word repetitions, and character n-gram repetitions. This post-processing step reduces WER by cleaning artefacts from the model's output without affecting valid transcriptions.

\subsection{Diarization Analysis}

\subsubsection{Value of Domain-Specific Fine-Tuning}
The pretrained PyAnnote segmentation model, while multilingual, was not specifically trained on Bangla conversational speech. Fine-tuning on the competition's annotated data allowed the model to learn Bangla-specific speaker turn patterns, prosodic cues, and acoustic characteristics. The improvement in DER validates the importance of domain adaptation even when starting from a strong pretrained model.

\subsubsection{Segmentation-Only Fine-Tuning Strategy}
Our design choice to fine-tune \textit{only} the segmentation component and retain the pretrained speaker embedding and clustering modules proved effective. The segmentation model, which determines when'' speakers are active, benefits most from domain-specific data. The speaker embedding model, which captures what'' each speaker sounds like, generalizes well across languages since it operates on acoustic features rather than linguistic content.

\subsubsection{Post-Processing Impact}
The minimum-duration filter (0.3,s threshold) effectively removes noise-induced false speaker segments. Ultra-short segments are typically produced when the segmentation model briefly misclassifies noise or silence as speech.

\subsection{Challenges and Limitations}

\begin{itemize}
    \item \textbf{Computational constraints:} Due to limited GPU resources, full-weight fine-tuning was not performed, although it could potentially improve representation learning under the current task setting.

    \item \textbf{Limited diarization data:} The small number of annotated diarization files limits the fine-tuning capacity. The 2-file development split offers limited evaluation reliability.
    \item \textbf{Audio quality:} Some competition audio files exhibit severe noise and distortion that even augmented training cannot fully address.
\end{itemize}

\section{Conclusion}

We presented a comprehensive system for Bangla long-form ASR and speaker diarization as part of DL Sprint 4.0. For ASR, our fine-tuning of Whisper-Medium with extensive domain-specific augmentation and Bangla text normalization demonstrated significant WER reduction over the pretrained baseline. For diarization, selective fine-tuning of the PyAnnote segmentation model with a pipeline-swap strategy achieved notable DER improvements while maintaining computational efficiency. Key takeaways include: (1) full-weight fine-tuning with 8-bit optimizers is viable on consumer GPUs for medium-scale Whisper models; (2) segment-wise augmentation with diverse acoustic effects improves real-world ASR robustness; (3) fine-tuning only the segmentation component of a multi-stage diarization pipeline is an effective and practical strategy; and (4) Bangla-specific text normalization is essential for reducing training-label noise.

Future work will focus on improving efficiency and robustness for real world deployment by developing joint ASR diarization architectures with pyannote.audio aligned modules to enhance speaker assignment using transcription information. We also plan to apply model compression techniques such as quantization and pruning to reduce latency and memory usage, enabling deployment on resource constrained edge devices.


\end{document}